\documentclass[12pt,a4paper,twoside]{article}
\usepackage{latexsym,amssymb,bbm,graphicx,epsfig,here}
\usepackage[dvips]{rotating}
\newcommand{\nwc}{\newcommand}
\nwc{\be}  {\begin{equation}}
\nwc{\ee}  {\end{equation}}
\nwc{\ba}  {\begin{array}}
\nwc{\ea}  {\end{array}}
\nwc{\bdm} {\begin{displaymath}}
\nwc{\edm} {\end{displaymath}}
\nwc{\bea} {\begin{eqnarray}}
\nwc{\eea} {\end{eqnarray}}
\nwc{\bear} {\begin{eqnarray}}
\nwc{\ear} {\end{eqnarray}}
\nwc{\LL} {\cal L} 
\renewcommand{\bar}{\overline}

\begin{document}
\begin{titlepage}
\begin{flushright}
HD-THEP-00-27
\end{flushright}
\vspace{2cm}
\begin{center}
{\Large Spontaneous symmetry breaking in the colored Hubbard model}\\
\vspace{2cm}
Tobias Baier\footnote{e-mail: baier@thphys.uni-heidelberg.de}, Eike Bick\footnote{e-mail: bick@thphys.uni-heidelberg.de}, Christof Wetterich\footnote{e-mail: C.Wetterich@thphys.uni-heidelberg.de}\\
\bigskip
Institut  f\"ur Theoretische Physik\\
Universit\"at Heidelberg\\
Philosophenweg 16, D-69120 Heidelberg\\
\vspace{3cm}
\end{center}

\begin{abstract}

The Hubbard model is reformulated in terms of different ``colored'' fermion species for the electrons or holes at different lattice sites. Antiferromagnetic ordering or d--wave superconductivity can then be described in terms of translationally invariant expectation values for colored composite scalar fields. A suitable mean field approximation for the two dimensional colored Hubbard model shows indeed phases with antiferromagnetic ordering or d--wave superconductivity at low temperature. At low enough temperature the transition to the antiferromagnetic phase is of first order. The present formulation also allows an easy extension to more complicated microscopic interactions.
\end{abstract}
\end{titlepage}

The Hubbard model \cite{hubbard} is one of the most studied models for electron systems. In particular, the two dimensional model appears to be a good candidate \cite{supra2d} for an explanation of high $T_c$ superconductivity. Despite its simplicity, several obstacles render even its approximate solution a difficult theoretical task. As a fermionic system it is not easily accessible to numerical simulations. Furthermore, there seems to be a competition between antiferromagnetic order and d--wave superconductivity. Both operators do not correspond to fermion bilinears on the same lattice site.

Recently, promising investigations using approximations to exact renormalization group equations \cite{RG} have been started \cite{Sal97}, \cite{Met99}. The difficulty of these approaches, however, consists in the high complexity of the equations if the full momentum dependence of correlation functions for several fermions is kept. In particular, the low temperature phases can only be realistically described if effective interactions involving more than four fermions are included. In our opinion a prerequisite for a successful use of these methods in the ordered phases is a simplification of the momentum dependence of the interactions. This can be done if the most prominent physical degrees of freedom are identified. We propose here a version of the Hubbard model where the relevant order parameters correspond to translationally invariant vacuum expectation values of scalar fields. We will see that this formulation can describe the low temperature phases in a very simple way. We therefore hope that it constitutes a good starting point for a detailed renormalization group analysis.

We consider the partition function of the Hubbard model \cite{negele}
\be
\label{1}
Z=\int D\hat\psi D\hat\psi^*\exp\Big\{-S[\hat\psi,\hat\psi^*]
+\int^\beta_0 d\tau\sum_i(\eta^*_i\hat\psi_i+\eta_i\hat\psi^*_i)-\tilde S_j\Big\}
\ee
\bea
\label{2}
S[\hat\psi,\hat\psi^*]&=&\int^\beta_0d\tau\Big\{\sum_i\Big(\hat\psi^*_i
  \partial_\tau\hat\psi_i - \mu\hat\psi^*_i\hat\psi_i\nonumber\\
&&-\frac{1}{6}U(\hat\psi^*_i\vec\tau\hat\psi_i)
  (\hat\psi^*_i \vec\tau\hat\psi_i)\Big) + \sum_{ij}\hat\psi^*_i {\cal T}_{ij}\hat\psi_j\Big\}
\eea
as a functional of sources $\eta, \eta^*$ of the fermions as well as  
sources of fermion bilinears ($\tilde S_j$) that will be
specified below (cf. eq. (\ref{9})). The spinors $\hat\psi_i=(\hat\psi_{i\uparrow},\hat\psi_{i\downarrow})^T$, $\hat\psi_i^*=(\hat\psi^*_{i\uparrow},\hat\psi^*_{i\downarrow})$ (as well as the fermionic sources $\eta_i$, $\eta^*_i$) are two-component Grassmann variables depending on the Euclidean ``time'' $\tau$ with
antiperiodicity $\hat\psi_i(\beta)=-\hat\psi_i(0),\ \hat\psi_i^*(\beta)=-\hat\psi^*_i(0)$. Here $\beta=1/T$ is the inverse temperature. 
We treat $\psi$ and $\psi^*$ as independent Grassmann variables, even though
the notation is reminiscent of a type of complex conjugation which also inverts $\tau$
and reorders all Grassmann variables. (In quantum field theory the invariance of the 
action under this discrete transformation is related to Osterwalder-Schrader
positivity.) 

The index $i$ labels the sites of the lattice. We concentrate on a
quadratic lattice in two dimensions, $i=(m,n)$, $m, n\in {\mathbbm
  Z}$, with next neighbor interactions where ${\cal T}_{ij}=-t$ for $i,j$
next neighbors and ${\cal T}_{ij}=0$ otherwise. They describe the probability of fermion tunneling between different
lattice sites. After a Fourier transform the Fermi surface for $U\to 0$ is given by 
\bea
\label{3}
&&-2t \left( \cos(aq_1)+\cos(aq_2) \right) =\mu\nonumber\\
&&|q_\mu|\leq 2\Lambda,\ \Lambda=\pi/2a 
\eea
with the lattice spacing $a$. The local\footnote{Here the term local refers to our neglect of the interaction 
of fermions located at {\em different} lattice sites.} Coulomb interaction of the fermions
involves the Pauli matrices $\vec\tau$ and can be rearranged, e.g.
$(\hat\psi^*_i\vec\tau\hat\psi_i)(\hat\psi^*_i\vec\tau\hat\psi_i)
=-3(\hat\psi_i^*\hat\psi_i)(\hat\psi_i^*\hat\psi_i) =-6 n_{i\uparrow}
n_{i\downarrow}$ with $n_{i\uparrow (\downarrow)} =
\hat\psi^*_{i\uparrow(\downarrow)} \hat\psi_{i\uparrow(\downarrow)}$. As
usual, the expectation values of operators are related to appropriate
derivatives of $Z$ with respect to the sources.
In addition to the discrete lattice symmetries, the model has two obvious continuous symmetries: the $SU(2)$-spin rotations and the $U(1)$-phase rotations corresponding to charge conservation.

In units where $\hbar=c=k_B=1$, the parameters $T$, $\mu$, $t$
and $U$ have dimension of mass whereas $\hat\psi, \hat\psi^*$ are
dimensionless. The partition function is invariant under the rescaling
$(\alpha\in {\mathbbm R}_+)$ $\tau\to\tau/\alpha, \mu\to\alpha\mu, t\to\alpha t,
U\to\alpha U, T\to\alpha T$.  It can therefore only
depend\footnote{This holds up to a possible temperature-dependent
  factor from the functional measure.} on the dimensionless parameter
ratios $\mu/U, T/U, t/U$ and is independent of the dimensionless combination $aU$. 
Furthermore, the invariance of $Z$ under the discrete transformation
$\hat\psi(\tau)\to-\hat\psi(\beta-\tau),\ 
\hat\psi^*(\tau)\to\hat\psi^*(\beta-\tau),\ \mu\to -\mu,\ t\to-t$
(with appropriate transformations of the sources) permits a
restriction to $\mu\geq0$. Finally, we may divide the lattice sites
$i$ into two classes, $i\in I_1$, if $m$ and $n$ are both even or both
odd, $i\in I_2$ otherwise. The transformation 
$\hat\psi_{i\in I_2}\to-\hat\psi_{i\in I_2}$, 
$\hat\psi^*_{i\in I_2}\to-\hat\psi^*_{i\in I_2}$ while leaving 
$\hat\psi_{i\in I_1},\hat\psi^*_{i\in I_1}$ invariant maps $Z(t)\to Z(-t)$ (again
with an appropriate mapping for the sources). 
We restrict our discussion to positive $t$ and $\mu$. Predictions for models with negative $t$ or negative $\mu$
can easily be obtained from our results by an appropriate mapping.

\begin{figure}[th]
  \begin{center}
  \unitlength1mm
  \includegraphics[scale=1.0]{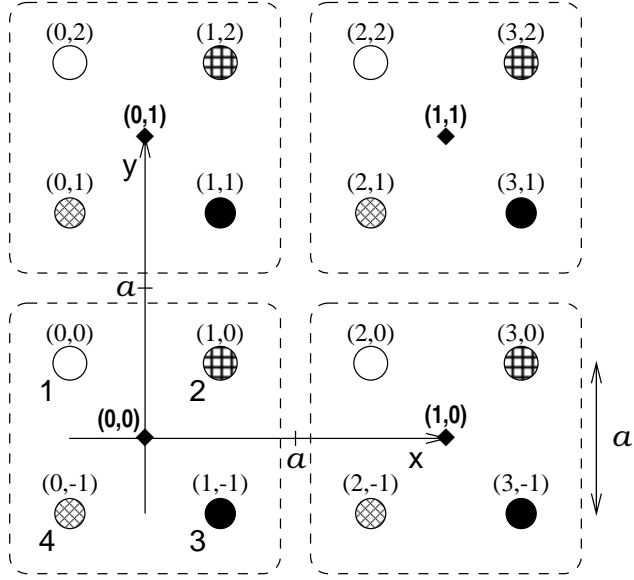}
  \caption{The colored Hubbard model. The lattice sites of the coarse lattice are symbolized by a $\blacklozenge$.
           The numbering of the sites of our original lattice is also shown.}
  \label{fig:colormodel}
  \end{center}
\end{figure}
In order to represent the fermion bilinears of interest in a simple
form, we choose to label the variables at four neighboring lattice
sites by different colors red, yellow, green and blue,
$\hat\psi_a,\ a=1,...,4$. (This also allows us to easily extend the 
formalism to lattices with different atoms.)
We take $\vec x=(x,y)=(ma,na)$
with $m, n$ even as the sites of a coarse lattice with
lattice distance $2a$ and define 
\bea\label{4}
\hat\psi_1(\vec x)=\hat\psi_{m,n} \quad &,&\hat\psi_2(\vec x)=
\hat\psi_{m+1,n}\nonumber\\
\hat\psi_4(\vec x)=\hat\psi_{m,n-1}&,&\hat\psi_3(\vec x)=\hat\psi_{m+1,n-1}
\eea
and similar for $\hat\psi_a^*$ (see fig.\ref{fig:colormodel}).

The lattice symmetries $T_x$ (translation in $x$ by $a$),
$T_y$ (translation in $y$ by $a$), $R$ (clockwise rotation by
$90^\circ$ around $\vec x=0$), $I_x$ (reflection at $x$-axis)
and $\tilde I_x$ (reflection at the axis $y=\frac{1}{2}a$) act as
\bea\label{5}
T_x&:&\hat\psi_1(x,y)\to\hat\psi_2(x,y)\ ,\ \hat\psi_2(x,y)\to \hat\psi_1(x+2a,y)\nonumber\\
&&\hat\psi_4(x,y)\to\hat\psi_3(x,y)\ ,\ \hat\psi_3(x,y)\to \hat\psi_4(x+2a,y)\nonumber\\
T_y&:& \hat \psi_1(x,y)\to\hat\psi_4(x,y+2a)\ ,\ \hat\psi_2(x,y)\to\hat\psi_3(x,y+2a)\nonumber\\
  &&\hat\psi_3(x,y)\to\hat\psi_2(x,y)\ ,\ \hat\psi_4(x,y)\to\hat\psi_1(x,y)\nonumber\\
R&:&\hat\psi_1(x,y)\to\hat\psi_2(y,-x)\ ,\ \hat\psi_2(x,y)\to \hat\psi_3(y,-x)\nonumber\\
  &&\hat\psi_3(x,y)\to\hat\psi_4(y,-x)\ ,\ \hat\psi_4(x,y)\to \hat\psi_1(y,-x)\nonumber\\
I_x&:&\hat\psi_1(x,y)\leftrightarrow \hat\psi_4(x,-y)\ ,\ 
  \hat\psi_2(x,y)\leftrightarrow\hat\psi_3(x,-y)\nonumber\\
\tilde I_x&:&\hat\psi_{1,2}(x,y)\to \hat\psi_{1,2}(x,-y)\ ,\ 
  \hat\psi_{3,4}(x,y)\to\hat\psi_{3,4}(x,-y+2a).
\eea
The lattice symmetries of the coarse lattice can be composed 
from the generators $T_x,R,I_x$. We note that they also act in 
color space\footnotemark. 

\footnotetext{The local interaction is also invariant under relabelings
at fixed $\vec x$
\begin{eqnarray*}
r&:&\hat\psi_1\to\hat\psi_2\to\hat\psi_3\to\hat\psi_4\nonumber\\
i&:&\hat\psi_1\leftrightarrow\hat\psi_4\ ,\ \hat\psi_2
\leftrightarrow\hat\psi_3\nonumber\\
d&:&\hat\psi_1\leftrightarrow\hat\psi_3\ ,\ \hat\psi_2\leftrightarrow
\hat\psi_4.
\end{eqnarray*}
This, however, is not a symmetry of the next neighbor interaction.}

Useful fermion bilinear operators are $(c=i\tau_2)$
\bea\label{7}
\tilde\sigma_{ab}(\vec x)&=&\hat\psi_b^*(\vec x)\hat\psi_a(\vec x)\nonumber\\
\vec{\tilde \varphi}_{ab}(\vec x)&=&\hat\psi^*_b(\vec x) \; \vec\tau \; \hat\psi_a(\vec x)\nonumber\\
\tilde\chi_{ab}^{(1)}(\vec x)&=&\hat\psi^T_b(\vec x) \; c \; \hat\psi_a(\vec x)
  =\tilde\chi^{(1)}_{ba}(\vec x)\nonumber\\
\tilde\chi_{ab}^{(2)}(\vec x)&=&-\hat\psi^*_b(\vec x) \; c \; \hat\psi^{*T}_a(\vec x)
  =\tilde\chi^{(2)}_{ba}(\vec x).
\eea
(We have omitted for simplicity fermion-fermion pairs in the triplet of the spin
group, similar to the fermion-antifermion pairs $\vec{\tilde\varphi}$. They are antisymmetric
in the color indices.) Among the electron-electron (or hole-hole)
pairs we concentrate on
\bea\label{7a}
\tilde s^{(\alpha)}&=&\tilde\chi_{11}^{(\alpha)}+
 \tilde\chi_{22}^{(\alpha)}+\tilde\chi_{33}^{(\alpha)}+\tilde\chi^{(\alpha)}_{44}\nonumber\\
\tilde c^{(\alpha)}&=&\tilde\chi_{11}^{(\alpha)}-
 \tilde\chi_{22}^{(\alpha)}+\tilde\chi_{33}^{(\alpha)}-\tilde\chi^{(\alpha)}_{44}\nonumber\\
\tilde d^{(\alpha)}&=&\tilde\chi_{12}^{(\alpha)}-
 \tilde\chi_{23}^{(\alpha)}+\tilde\chi_{34}^{(\alpha)}-\tilde\chi^{(\alpha)}_{41}\nonumber\\
\tilde e^{(\alpha)}&=&\tilde\chi_{12}^{(\alpha)}+
 \tilde\chi_{23}^{(\alpha)}+\tilde\chi_{34}^{(\alpha)}+\tilde\chi^{(\alpha)}_{41}\nonumber\\
\tilde v^{(\alpha)}_x &=&\tilde\chi_{23}^{(\alpha)}-\tilde\chi_{41}^{(\alpha)}\ ,\ 
 \tilde v_y^{(\alpha)} \;=\; \tilde\chi_{12}^{(\alpha)}-\tilde\chi^{(\alpha)}_{34}\nonumber\\
\tilde t^{(\alpha)}_1 &=&\tilde\chi_{11}^{(\alpha)}-\tilde\chi_{33}^{(\alpha)}\ ,\ 
 \tilde t_2^{(\alpha)} \;=\; \tilde\chi^{(\alpha)}_{22}-\tilde\chi^{(\alpha)}_{44}.
\eea
which transform as $s$-wave $(\tilde s)$, $d_{xy}$-wave
$(\tilde c), d_{x^2-y^2}$-wave $(\tilde d)$, extended $s$-wave
$(\tilde e)$, $p$-wave $(\tilde v_x,\tilde v_y)$ in the spin singlet
state. Similarly, for the electron-hole pairs we select
\bea\label{8}
\tilde\rho&=&\tilde\sigma_{11}+\tilde\sigma_{22}+\tilde\sigma_{33}+
\tilde\sigma_{44}\nonumber\\
\tilde p&=&\tilde\sigma_{11}-\tilde\sigma_{22}+\tilde\sigma_{33}-
\tilde\sigma_{44}\nonumber\\
\tilde q_1&=&\tilde\sigma_{11}-\tilde\sigma_{33}\ ,\ 
  \tilde q_2 \;=\; \tilde\sigma_{22}-\tilde\sigma_{44}\nonumber\\
\vec{\tilde m}&=&\vec{\tilde \varphi}_{11}+\vec{\tilde\varphi}_{22}+\vec
{\tilde\varphi}_{33}+\vec{\tilde\varphi}_{44}\nonumber\\
\vec{\tilde a}&=&\vec{\tilde \varphi}_{11}-\vec{\tilde\varphi}_{22}+\vec
{\tilde\varphi}_{33}-\vec{\tilde\varphi}_{44}\nonumber\\
\vec{\tilde g}_+&=&\vec{\tilde\varphi}_{11}-\vec{\tilde\varphi}_{33}\ , \ 
  \vec{\tilde g}_- \;=\; \vec{\tilde\varphi}_{22}-\vec{\tilde\varphi}_{44}
\eea
They correspond to the charge density $\tilde\rho$, the charge
modulation or charge density wave $\tilde p$, the ferromagnetic and antiferromagnetic
spin densities $\vec{\tilde m}, \vec{\tilde a}$ and the
diagonal spin density $\vec{\tilde g}_\pm$. 
Correspondingly, we specify the source term for the bilinears as
\bea\label{9}
\tilde S_j&=&-\int d\tau\sum_{\vec x}\Big(\sum_{\beta}
\{j^*_\beta(\vec x)\tilde u^{(1)}_\beta(\vec x)+j_\beta(\vec x)\tilde u^{(2)}_\beta(\vec x)
  +\tilde r_\beta j^*_\beta(\vec x)j_\beta(\vec x) \}\\
&&\qquad 
  +\sum_\gamma \{l_\gamma(\vec x)\tilde w_\gamma(\vec x)+\frac{1}{2}r_\gamma l_\gamma(\vec x) l_\gamma(\vec x)\}
-\mu\tilde w_\rho(\vec x)\Big)
  +\frac{1}{8}r_\rho \beta V_2\mu^2/a^2\nonumber
\eea
with 
\bea 
\tilde u^{(\alpha)}&=&(\tilde s, \tilde c,\tilde d, \tilde e,\tilde v_x, \tilde v_y, \tilde t_1, \tilde t_2)^{(\alpha)}, \nonumber\\
\tilde w &=& (\tilde \rho, \tilde p, \tilde q_1,\tilde q_2 ,\vec{\tilde m}, \vec{\tilde a}, \vec{\tilde g}_+, \vec{\tilde g}_-). \nonumber
\eea 
The complex sources $j=(j_s,j_c, j_d,j_e,j_{v_x},j_{v_y},j_{t_1},j_{t_2})$ 
and the real sources 
$l=(\mu+l_\rho',l_p,l_{q_1}, l_{q_2}, \vec l_m,\vec l_a, \vec l_{g_+}, \vec l_{g_-})$ 
also depend on $\tau$ and from now on we include the chemical
potential $\mu$ in the  source for $\tilde \rho$. Here $V_2$ 
is the two-dimensional volume and we will specify the constants $\tilde r_\beta$
and $r_\gamma$ below. 

Using the identities
\bea\label{10}
\tilde\chi^{(2)}_{ab}\tilde\chi^{(1)}_{cd}&=&\tilde\sigma_{ca}\tilde
  \sigma_{db}+\tilde\sigma_{cb}\tilde\sigma_{da}\nonumber\\
\vec{\tilde\varphi}_{ab}\vec{\tilde\varphi}_{cd}&=&-\tilde\sigma_{ab} \tilde\sigma_{cd}
  -2\tilde\sigma_{ad}\tilde\sigma_{cb}
\eea
we establish the relations
\begin{eqnarray}\label{11}
\frac{1}{2}\left(\tilde{s}^{(2)}\tilde{s}^{(1)}+\tilde{c}^{(2)}\tilde{c}^{(1)}\right)
  +\tilde{t}_1^{(2)}\tilde{t}_1^{(1)}+\tilde{t}_2^{(2)}\tilde{t}_2^{(1)}
  &=& 4 \sum_a\tilde{\sigma}_{aa}^2 \nonumber\\ 
\frac{1}{2}\left(\tilde{\rho}^2+\tilde{p}^2\right)+\tilde{q}_1^2+\tilde{q}_2^2
  &=& 2 \sum_a\tilde{\sigma}_{aa}^2 \nonumber\\
\frac{1}{2}\left(\vec{\tilde{m}}^2-\vec{\tilde{a}}^2-\tilde{\rho}^2+\tilde{p}^2\right)
  +\tilde{d}^{(2)}\tilde{d}^{(1)}+\tilde{e}^{(2)}\tilde{e}^{(1)} \hspace{1cm} && \nonumber\\
{}+ 2\left(\tilde{v}_x^{(2)}\tilde{v}_x^{(1)}+\tilde{v}_y^{(2)}\tilde{v}_y^{(1)}\right)
  &=& 0 \nonumber\\
\frac{1}{2}\left(\vec{\tilde{a}}^2+\vec{\tilde{m}}^2\right)
  +\vec{\tilde{g}}_+^2+\vec{\tilde{g}}_-^2
  &=& -6 \sum_a\tilde{\sigma}_{aa}^2.
\end{eqnarray}
The interaction term in (\ref{2}) reads $\frac{U}{2}\sum_a\tilde{\sigma}_{aa}^2$ and it is obvious that the
decomposition into products of the above fermion bilinears is not unique.

It is now straightforward to derive a partially  bosonized version of the
Hubbard model by introducing a suitable identity in the functional
integral (\ref{1}). Inversely, the bosonized partition function 
\bea\label{12}
Z&=&\exp\left\{-2\frac{\Lambda^2}{h^2_\rho}\beta V_2\mu^2 \right\}
\int D\hat\psi D\hat\psi^*D\hat u \, D\hat u^* D\hat w \nonumber\\
  &&\qquad\qquad\qquad \exp\Big\{-\int d\tau \sum_{\vec x}({\cal L}_{kin}+{\cal L}_Y+{\cal L}_B+{\cal L}_j)\Big\}
  \nonumber\\
{\cal L}_{kin}&=&\sum_a\hat\psi_a^*(\vec x) \;\partial_\tau\; \hat\psi_a(\vec x)\nonumber\\
&&{}-t\big\{\hat\psi_1^*(x,y)[\hat\psi_2(x,y)+\hat\psi_2(x-2a,y)+\hat\psi_4(x,y+2a)+\hat\psi_4(x,y)]\nonumber\\
&&\quad{}+\hat\psi_2^*(x,y)[\hat\psi_1(x+2a,y)+\hat\psi_1(x,y)+\hat\psi_3(x,y+2a)+\hat\psi_3(x,y)]\nonumber\\
&&\quad{}+\hat\psi_3^*(x,y)[\hat\psi_4(x+2a,y)+\hat\psi_4(x,y)+\hat\psi_2(x,y)+\hat\psi_2(x,y-2a)]\nonumber\\
&&\quad{}+\hat\psi_4^*(x,y)[\hat\psi_3(x,y)+\hat\psi_3(x-2a,y)+\hat\psi_1(x,y)+\hat\psi_1(x,y-2a)]\big\}\nonumber\\
{\cal L}_B &=& 4\pi^2 \sum_\beta \hat u_\beta^*(\vec x)\hat u_\beta
  (\vec x)+2\pi^2\sum_\gamma \hat w_\gamma(\vec x) \hat w_\gamma(\vec x)
  \nonumber\\
{\cal L}_Y&=&-\sum_\beta\tilde h_\beta(\hat u_\beta^*(\vec x)\tilde u_\beta^{(1)}(\vec x)+
  \hat u_\beta(\vec x)\tilde u_\beta^{(2)}(\vec x)) 
  -\sum_\gamma h_\gamma\hat w_\gamma(\vec x)\tilde w_\gamma(\vec x) \nonumber\\
{\cal L}_j&=&-\sum_\beta\frac{4\pi^2}{\tilde h_\beta}
  (j^*_\beta(\vec x)\hat u_\beta(\vec x)+j_\beta(\vec x)\hat u_\beta^*(\vec x))-
  \sum_\gamma\frac{4\pi^2}{h_\gamma} l_\gamma(\vec x) \hat w_\gamma(\vec x)\nonumber\\
&&-\sum_a\left(\eta_a^*(\vec x)\hat\psi_a(\vec x)+\eta_a(\vec x)\hat\psi_a^*(\vec x)\right) 
\eea
can easily be transformed into a purely fermionic functional integral
by performing the Gaussian integration over the complex scalar fields
$\hat u$ and real scalar fields $\hat w$. 
We choose $\tilde r_\beta=4\pi^2/\tilde h_\beta^2, \; r_\gamma=4\pi^2/h^2_\gamma$ such that the
partition functions (\ref{1}) and (\ref{12}) coincide
except for the quartic interactions. Indeed, the four fermion
interaction resulting from the bosonic functional integration
can be more complex than in the original Hubbard model, i.e.
\be
\label{13}
{\cal L}_{int}=-\sum_\beta\frac{\tilde h_\beta^2}{4\pi^2}
\tilde u^{(2)}_\beta\tilde u^{(1)}_\beta-\sum_\gamma
\frac{h^2_\gamma}{8\pi^2}\tilde w_\gamma\tilde w_\gamma.
\ee
Only for particular values of the real positive Yukawa couplings
$\tilde h_\beta,h_\gamma$ the partition function (\ref{12}) is equal to the partition
function (\ref{1}) of the Hubbard model\footnotemark, namely for (cf. eq.(\ref{11}))
\footnotetext{This holds up to an
irrelevant source-independent overall normalization factor.}

\begin{eqnarray}\label{14}
&&\tilde{h}_\beta^2 = \frac{\pi^2}{3}\tilde{H}_\beta U,
  \quad h_\gamma^2=\frac{\pi^2}{3}H_\gamma U \nonumber\\
&&2\tilde{H}_s = 2\tilde{H}_c=\tilde{H}_{t_1}=\tilde{H}_{t_2}=3\lambda_1 \nonumber\\
&&2\tilde{H}_d = 2\tilde{H}_e=\tilde{H}_{v_x}=\tilde{H}_{v_y}=6\lambda_3 \nonumber\\
&&H_{q_1} = H_{q_2}=6\lambda_2 \nonumber\\
&&H_\rho = 3\left(\lambda_2-\lambda_3\right), 
  \quad H_p=3\left(\lambda_2+\lambda_3\right) \nonumber\\
&&H_{\vec{a}} = 2\lambda_1+\lambda_2-3\lambda_3+1 \nonumber\\
&&H_{\vec{m}} = 2\lambda_1+\lambda_2+3\lambda_3+1 \nonumber\\
&&H_{\vec{g}_+} = H_{\vec{g}_-}=4\lambda_1+2\lambda_2+2,
\end{eqnarray}
where the parameters $\lambda_i$ obey
\begin{eqnarray*}
\lambda_i & > & 0\qquad \forall i=1\ldots3,\\
\lambda_2 & > & \lambda_3,\\
2\lambda_1+\lambda_2+1 & > & 3\lambda_3.
\end{eqnarray*}

We emphasize that the choice (\ref{14}) of the Yukawa couplings is not unique
since it depends on the three parameters $\lambda_i$. Arbitrary values of $\lambda_i$
(within the allowed range) all describe the same Hubbard model. The independence of 
physical results on the values of $\lambda_i$ can be used 
as a check for the validity of approximations.
Furthermore, a large variety of different four-fermion interactions can be described by varying the
Yukawa couplings away from the ``Hubbard values'' (\ref{14}).

The symmetries $R$ and $I_x$ as well as the translations by $2a$ are easily
implemented on the space of bilinears $\tilde u_\beta,\tilde w_\gamma$
and correspondingly for the scalar fields $\hat u_\beta,\hat w_\gamma$.
This is not the case for the translations by $a$. The above 
formulation of the bosonization is therefore not optimal yet
if -- beyond the symmetries of the coarse-grained lattice -- the
symmetries like $T_x$ play an important role (as for the Hubbard
model). It is easy to remedy this shortcoming by an extension of
the space of bilinears and the corresponding scalar  fields.
We introduce an additional color index for the fermion bilinears
and the scalars by
\bea\label{15}
&&\tilde w_{1\gamma}(\vec x)=T_yT_x^{-1}\tilde w_\gamma(\vec x) \ ,
  \ \tilde w_{2\gamma}(\vec x)=T_y\tilde w_\gamma(\vec x)\ ,
  \nonumber\\
&&\tilde w_{3\gamma}(\vec x)=\tilde w_\gamma (\vec x)
  \ ,\ \tilde w_{4\gamma}(\vec x)=T_x^{-1}\tilde w_\gamma(\vec x)
\eea
and similar for $\tilde u^{(\alpha)}, \hat w, \hat u^{(\alpha)}, l, j$. Products
like $w_\gamma w_\gamma$ are now understood as scalar products
\be\label{16}
w_\gamma w_\gamma\to\frac{1}{4}\sum_a w_{a\gamma}w_{a\gamma}.
\ee
With these replacements\footnote{The term $\sim\mu\tilde w_\rho(\vec x)$ in eq. (\ref{9}) becomes $(\mu/4)\sum_a\tilde w_{a\rho}(\vec x)$.} it is straightforward to check that the
partition function (\ref{12}) with the choice of Yukawa couplings 
(\ref{14}) is again exactly equal to the one of the Hubbard model
if all sources except $\mu$ are zero. The translations $T_x,T_y$ are
now directly implemented on the scalar fields, e.g.
$T_x(\hat w_{1\gamma}(\vec x))=\hat w_{2\gamma}(\vec x)$. The same holds
true for the rotation $\tilde R$ or the reflection $\tilde I_x$.

In conclusion, we have developed an extended version of the Hubbard model
-- the colored Hubbard model -- which coincides with the Hubbard model
for special values of the Yukawa couplings (\ref{14}) and the 
sources. Particularly simple modifications arise for $\vec x$-independent
and $\tau$-independent sources. As an example, the source
\be\label{17}
l_{3\rho}' = l_{3p} = -2 l_{3q_1} = -\frac{1}{4}\nu
\ee
induces an additional energy for the occupation of the sites $(m,n)$
with both $m$ and $n$ odd
\be\label{18}
S_\nu=\nu\int d\tau\sum_{{m\ \mathrm{odd}}\atop{n\ \mathrm{odd}}}\hat\psi_{mn}^* \hat\psi_{mn}.
\ee
For $\nu\to\infty$ these sites are effectively removed from the
lattice and we therefore deal with the Hubbard model on a non-quadratic
lattice structure.

Analytic computations for the partition function (\ref{12}) are most easily done in momentum space.
It is straightforward to perform a Fourier transform using
\bea\label{Y1}
\hat\psi_a(\vec x,\tau) &=& \sqrt{2a}T\sum_n\int\frac{d^2q}{(2\pi)^2}
\exp \left(i\{(\vec x+\vec z_a)\vec q+2\pi nT\tau\}\right)\hat\psi_{an}(\vec q)\nonumber\\
\hat\psi_a^*(\vec x,\tau) &=& i\sqrt{2a} T\sum_n\int\frac{d^2q}{(2\pi)^2}
\exp\left(-i\{(\vec x+\vec z_a)\vec q+2\pi nT\tau\}\right)\hat{\bar\psi}_{an}(\vec q)\gamma^0\nonumber\\
\hat\psi^* &=& i\hat{\bar \psi}\gamma^0\ ,\ 
  \gamma^0=\left(\begin{array}{cc}\tau_3&0\\0&\tau_3\end{array}\right).
\eea
Here the Matsubara frequencies are labeled by half integer $n = \pm 1/2, \pm 3/2,\ldots)$ and
the momentum integration is in the range $-\Lambda<q_x<\Lambda, \; -\Lambda<q_y<\Lambda$ as
appropriate for the coarse lattice with lattice distance $2a$. We choose
\begin{eqnarray}
\label{Y12}
&&\vec z_1=\left(-\frac{a}{2},\frac{a}{2}\right) \ , \ \vec z_2=\left(\frac{a}{2}\ ,\ \frac{a}{2}\right), \nonumber\\
&&\vec z_3=\left(\frac{a}{2}\ ,\ -\frac{a}{2}\right) \ , \ \vec z_4=\left(-\frac{a}{2}\ ,\ -\frac{a}{2}\right)
\end{eqnarray}
corresponding to an expansion in the coordinates of the
$(m,n)$ lattice. This yields for the kinetic term
\bea\label{Y2}
S_{kin}&=&\int d\tau\sum_{\vec x}{\cal L}_{kin} \nonumber\\
&=&T\sum_n\int\frac{d^2q}{(2\pi)^2}\hat{\bar\psi}_{an}(\vec q)P^{(0)}_{ab}(n,\vec q)\hat\psi_{bn}(\vec q)\nonumber\\
P^{(0)}&=&-2\pi nT\gamma^0-2it\gamma^0\{\cos(aq_x)A_1+\cos(aq_y)B_1\}
\eea
where we use matrices ($\tau_0 = {\mathbbm 1}_2$; $\mu=0 \ldots 3$; $i,j=1 \ldots 3$)
\bea\label{Y3}
&&A_\mu=\left(
\begin{array}{cc}
  \tau_\mu & 0\\ 0 &\tau_\mu
\end{array}\right)\ , \ 
B_\mu= \left(
\begin{array}{cc}
  0 & \tau_\mu\\ \tau_\mu & 0
\end{array}\right)\nonumber\\
&&\{A_i,B_j\}=2\delta_{ij} B_0\ ,\ \{A_i,A_j\} =
\{ B_i,B_j\}=2\delta_{ij}\nonumber\\
&&[A_i,B_j]=2i\epsilon_{ijk}B_k\nonumber\\
&&B_0 A_i=A_i B_0=B_i\ ,\ B_0 B_\mu=B_\mu B_0=A_\mu.
\eea

Spontaneous symmetry breaking with ``extended order parameters''
like the antiferromagnetic spin density $\sim(\vec {\tilde a}_1
-\vec {\tilde a}_2+\vec {\tilde a}_3-\vec {\tilde a}_4)$ or
$d$-wave superconductivity with order parameter $\sim(\tilde d_1
+\tilde d_2+\tilde d_3+\tilde d_4)$ can be directly investigated
in our formalism by looking for the minima of the effective
scalar potential. The notion of the effective potential is a very powerful concept since it describes
simultaneously situations with vanishing and nonvanishing sources, i.e. in addition to the Hubbard model
for arbitrary $\mu$ it also comprises many extended models. The effective potential corresponds to the effective
action for homogeneous colored scalar fields. We define the scalar expectation values
in the presence of sources\footnote{The variation with respect to
$l_{a\rho}$ is performed at fixed $\mu$.}
\bea\label{19}
&&u_{a\beta}(\vec x)=\frac{\Lambda^2}{\pi^2}\frac{\partial}{\partial J^*_{a\beta}(\vec x)}\ln Z
  = \; <\hat u_{a \beta}(\vec x)>\nonumber\\
&&u^*_{a\beta}(\vec x)=\frac{\Lambda^2}{\pi^2}\frac{\partial}{\partial J_{a\beta}(\vec x)}\ln Z
  = \; <\hat u^*_{a \beta}(\vec x)>\nonumber\\
&&w_{a\gamma}(\vec x)=\frac{\Lambda^2}{\pi^2}\frac{\partial}{\partial L_{a\gamma}(\vec x)}\ln Z
  = \; <\hat w_{a\gamma}(\vec x)>
\eea
with
\be\label{20}
J_{a\beta}=\frac{\Lambda^2}{\tilde h_\beta}j_{a\beta}\ ,\ 
L_{a\gamma} =\frac{\Lambda^2}{h_\gamma}l_{a\gamma}.
\ee
With the usual Legendre transform one obtains\footnote{We concentrate in the following on 
$\psi_a=\left<\hat\psi_a\right>=0$, $\psi_a^*=\left<\hat\psi_a^*\right>=0$} the effective action $\Gamma$
\bea\label{21}
\Gamma[u,w,\psi,\psi^*]&=&-\ln Z+\int d\tau\sum_{\vec x}\sum_a\Big\{
\frac{\pi^2}{\Lambda^2}\sum_\beta(J_{a\beta}^*u_{a\beta}+J_{a\beta}u^*_{a\beta})\nonumber\\
&&\qquad\qquad {}+\frac{\pi^2}{\Lambda^2}\sum_\gamma L_{a\gamma}w_{a\gamma}
  +\eta^*_a\psi_a - \psi^*_a \eta_a \Big\}
\eea
which obeys
\be\label{22}
\frac{\partial\Gamma}{\partial w_{a\gamma}}=
\frac{\pi^2}{\Lambda^2}L_{a\gamma}\qquad \mathrm{etc.}
\ee
Performing the derivatives (\ref{19}) in the fermionic functional
integral, we can directly relate the scalar expectation values 
to the expectation values of fermionic bilinears
\bea\label{21a}
u_{a\beta}&=&\frac{\tilde h_\beta}{4\pi^2} <\tilde u^{(1)}_{a\beta}> +
  \frac{1}{\Lambda^2}J_{a\beta}\nonumber\\
u_{a\beta}^*&=&\frac{\tilde h_\beta}{4\pi^2} <\tilde u^{(2)}_{a\beta}> +
  \frac{1}{\Lambda^2}J_{a\beta}^*\nonumber\\
w_{a\gamma}&=&\frac{ h_\gamma}{4\pi^2} <\tilde w_{a\gamma}> +
  \frac{1}{\Lambda^2}L_{a\gamma}.
\eea
In particular, if all sources except $L_{a\rho}(\vec x)=\Lambda^2 \mu/h_\rho$
vanish, the scalar $w_{a\rho}$ has contributions linear in the electron density
$n=<\tilde\rho>/4a^2$ and the chemical potential
\be\label{22a}
w_{a\rho}=\frac{h_\rho}{4\Lambda^2}n+\frac{\mu}{h_\rho}.
\ee

We will mainly be interested in homogenous expectation values
and therefore in the effective scalar potential which can be
obtained from $\Gamma$ for $\vec x$- and $\tau$-independent scalar
fields and vanishing fermion fields by
\be\label{23}
U_0=T\Gamma/V_2.
\ee
The ground state of the Hubbard model corresponds to the minimum
of $U_0$ with respect to all fields except $\rho=\frac{1}{4}
\sum_a w_{a\rho}$ given by eq. (\ref{22a}), where $\mu$ obeys
\be\label{24}
\mu=\frac{h_\rho}{4\Lambda^2}\frac{\partial U_0}{\partial \rho}.
\ee
We are interested in possible expectation values of scalars
different from $\rho$. Such a spontaneous symmetry breaking arises
if for some range of $\rho$ the minimum of $U_0$ (at fixed $\rho$)
occurs for a nonvanishing  scalar field. 

In this paper we compute the effective potential $U_0$ in the
``mean field'' approximation. This means that only the fermionic
part of the functional integral (\ref{12}) is performed in a
homogenous ``background'' $\hat u=u,\ \hat w=w$. This integral
is Gaussian, and we can write the mean field expression for $U_0$ as
\bea\label{25}
U_0&=&U_{cl}+\Delta U\nonumber\\
U_{cl}&=&\Lambda^2\sum_\beta\sum_au^*_{a\beta}u_{a\beta}+\frac{\Lambda^2}{2}
  \sum_\gamma\sum_aw_{a\gamma}w_{a\gamma}+\frac{2\Lambda^2}{h_\rho^2}\mu^2 \nonumber\\
\Delta U&=&-\frac{1}{2}T\sum_n\int\frac{d^2q}{(2\pi)^2}\ln \det P(n,\vec q).
\eea
Here $P(q)$ is a $16\times 16$ matrix (including spinor indices)
for the inverse fermion propagation in the presence of scalar
background fields. It is defined by the part of the action quadratic
in the fermion fields
\bea\label{27}
S_2 &=& \frac{1}{2}T\sum_n\int\frac{d^2q}{(2\pi)^2}\tilde\psi_{-n}^T(-\vec q)P(n,\vec q)\tilde\psi_n(\vec q) \\
\tilde\psi_n(\vec q) &=&
\left(\begin{array}{c}
\hat\psi_{a,n}(\vec q) \\ 
\hat{\bar\psi}_{a,-n}(-\vec q)
\end{array}\right)
\eea
and we find from eq. (\ref{Y2})
\be\label{26}
P=\left(\begin{array}{cc}0&-P_0^T(-n)\\ P_0(n)& 0\end{array}\right)
  +\tilde P(\rho,d,\vec a,...)
\ee
where the second term reflects the influence of the background
through the Yukawa couplings. We explore here the dependence of
the effective potential on the charge density $\rho$, $d$-wave
pair condensation $d$ and antiferromagnetic order parameter
$\vec a$. We therefore take
\bea\label{28}
&&w_{1\rho}=w_{2\rho}=w_{3\rho}=w_{4\rho}=\rho\nonumber\\
&&u_{1d}=u_{2d}=u_{3d}=u_{4d}=d\nonumber\\
&&\vec w_{1a}=-\vec w_{2a}=\vec w_{3a}=-\vec w_{4a}=\vec a
\eea
and find\footnote{The choice of $\gamma^0$ in (\ref{Y1}) is not crucial for this calculation -- any orthogonal $\gamma^0$ will do.}
\bea\label{29}
\tilde P&=&-ih_\rho \rho 
\left(\begin{array}{cc}
  0 & -{\gamma^0}^T\\ \gamma^0 & 0
\end{array}\right)
-ih_a\vec a
\left(\begin{array}{cc} 
  0 &-A_3{\gamma^0}^T\otimes\vec\tau^T\\
  \gamma^0 A_3\otimes\vec\tau &0
\end{array}\right) \nonumber\\
&&-h_d
\left(\begin{array}{c}
  d^*[\cos(aq_x)A_1-\cos(aq_y)B_1] \qquad\qquad\qquad 0\qquad \\
  \qquad 0 \qquad\qquad d\gamma^0[\cos(aq_x)A_1-\cos(aq_y)B_1]{\gamma^0}^T
\end{array}\right)
\otimes c \nonumber\\
&=& -\tilde P^T.
\eea
With $G=\mathrm{diag}({\mathbbm 1},-i{\gamma^0}^T)$ and $\underline P = G P G^T$
one obtains 
\bea\label{30}
\ln\det P&=&\ln\det \underline P = 
  \frac{1}{2}\ln\det\left( \underline{P}B_0\underline{P}^TB_0 \right)\nonumber\\
&=&\ln \mathrm{det}_8\{(2\pi nT)^2+[2t\cos(aq_x)A_1+2t\cos(aq_y)B_1+h_\rho\rho]^2 \nonumber\\
&& {}+h_a^2 \vec a\vec a + 2h_\rho h_a\rho\vec aA_3\otimes\vec\tau\nonumber\\
&& {}+h^2_d d^* d[\cos(aq_x)A_1-\cos(aq_y)B_1]^2 \}.
\eea
It is easy to see that $\Delta U$ depends only on the invariants
$\delta=d^*d,\ \alpha=\vec a\vec a$ and $\rho$. Up to an additive
($T$-dependent) constant one finds
\bea\label{31}
\Delta U&=&-\frac{1}{2} T\sum_n \int\frac{d^2q}{(2\pi)^2} \mathrm{tr}_8\ln\Big\{\mathbbm{1} \nonumber\\
&&{}+\Big( \left[2t\cos(aq_x)A_1+2t\cos(aq_y)B_1+h_\rho \rho+h_a\sqrt\alpha 
  A_3\otimes\tau_3\right]^2\nonumber\\
&&{}\qquad+h^2_d\delta\left[\cos(aq_x)A_1-\cos(aq_y)B_1\right]^2\Big)/(2\pi nT)^2\Big\}\\
U_{cl}&=&2\Lambda^2\alpha+4\Lambda^2\delta+2\Lambda^2\rho^2+\frac{2\Lambda^2\mu^2}{h_\rho^2}\nonumber,
\eea
and, evaluating the Matsubara sum and the trace, finally
\begin{eqnarray*}
U_0 &=& 2\Lambda^2\alpha + 4\Lambda^2\delta + \frac{2\Lambda^2\mu^2}{h^2_\rho}
  - 2T\int \frac{d^2q}{(2\pi)^2} \sum_{\varepsilon_i,\varepsilon_j\in\{-1,1\}} \\
&&\ln\cosh \left(\frac{1}{2T} \sqrt{\Big(h_\rho\rho+\varepsilon_i\sqrt{4t^2(c_x+\varepsilon_jc_y)^2+h_a^2\alpha}\Big)^2
    +h_d^2\delta(c_x-\varepsilon_jc_y)^2}\right)
\end{eqnarray*}
with $c_x=\cos(aq_x)$, $c_y=\cos(aq_y)$.     

For large temperature the fluctuation contribution $\Delta U$ 
is suppressed $\sim T^{-1}$. The minimum of $U_0$ therefore occurs
for all $\rho$ at $\alpha=0,\ \delta=0$. As $T$ is lowered, the
fluctuations tend to destabilize the ``symmetric minimum''.
In particular, the fluctuation contribution to the mass term
for $\vec a$ is negative for not too large $\rho^2$ and the
one for $d$ is negative for all $\rho$
\bea\label{32}
\Delta M^2_a &=& 2\frac{\partial}{\partial\alpha}\Delta U_{|\alpha=\delta=0} \nonumber\\
&=&-2h_a^2T\sum_n \int\frac{d^2q}{(2\pi)^2} 
  \mathrm{tr}_4\{P_\rho^{-2}-2h^2_\rho\rho^2 A_3P_\rho^{-2}A_3P_\rho^{-2}\}\nonumber\\
\Delta M^2_d&=&\frac{\partial}{\partial\delta}\Delta  U_{|\alpha=\delta=0} \nonumber\\
&=& -h^2_dT\sum_n \int\frac{d^2q}{(2\pi)^2} \mathrm{tr}_4\{(\cos(aq_x)A_1 - \cos(aq_y)B_1)^2P^{-2}_\rho\}
\eea
with
\be\label{33}
P^2_\rho=(2\pi nT)^2+(2t\cos(aq_x)A_1+2t\cos(aq_y) B_1+h_\rho\rho)^2
\ee
and $\mathrm{tr}_4$ the trace in color space only. These contributions
should be compared with $(M^{(0)}_a)^2=(M^{(0)}_d)^2=4\Lambda^2$.
The zeroes of $P_\rho^2$ for $T=0$ correspond to the
Fermi surface (\ref{3}) with shifted chemical potential
$\mu_{\mathrm{eff}}=h_\rho\rho$. (Neglecting contributions from $\Delta U$
eqs. (\ref{24}) and (\ref{25}) imply $\mu_{\mathrm{eff}}=\mu$.) We recall that
the momenta are restricted to the range corresponding to the coarse
lattice $|q_{x,y}|\leq\pi/(2a)=\Lambda$. On the
other hand we have now possible zeroes for different linear color
combinations. Noting that the eigenvalues of $A_1$ and $B_1$ are
$\pm1$, they precisely correspond to the original Fermi surface
-- the original zeros in the four ranges $|q_{x,y}|\leq\Lambda,\ \Lambda\leq
|q_{x,y}|\leq 2\Lambda,\ |q_x|\leq\Lambda,\ \Lambda\leq |q_y|\leq2\Lambda$ and
$\Lambda\leq |q_x| \leq2\Lambda,\ |q_y| \leq\Lambda$ appear now for
different color combinations in the range $|q_{x,y}|\leq\Lambda$.
Due to these zeros one finds $\lim_{T\to 0}\Delta M_d^2\to-\infty$
for not too large $\rho$ and similar for $\Delta M^2_a$ in the appropriate range of $\rho$.
This clearly indicates spontaneous symmetry breaking with $d$--wave
superconductivity or/and antiferromagnetic order parameter.
Note that in contrast to its derivatives the potential is not singular
for $T\to 0$. Since for large $\alpha$ and $\delta$ $U_0$ grows $\sim
(\alpha,\delta)$, the minimum occurs necessarily for finite 
$\alpha\not=0$ or $\delta\not=0$ if $M_a^2=4\Lambda^2+\Delta M_a^2$
or $M_d^2=4\Lambda^2+\Delta M_d^2$ become negative.

The spontaneous symmetry breaking cuts off the singularity near the Fermi surface or reduces its strength. An antiferromagnetic expectation value typically produces a gap for the fermionic fluctuations. For $\alpha>0,\;\delta=0,\; \rho=0$ this can be seen from a search for possible zeroes of $\det_8$ in eq. (\ref{30}) for $T=0$. On the other hand, for $\alpha=0,\; \delta>0$ the condition $\det_8 = 0$ requires $\cos(aq_x)=\pm\cos(aq_y)=\pm h_\rho\rho/(4t)$. In the superconducting phase the singularity therefore only occurs for special points in momentum space instead of a whole Fermi surface. As a consequence, the momentum integrations for the bosonic mass terms (similar to eq. (\ref{32})) remain finite even for $T\to 0$.

\begin{figure}[h]
\setlength{\unitlength}{0.8mm}

\begin{picture}(200,110)(0,-5)

\put(0,0){\epsfig{file=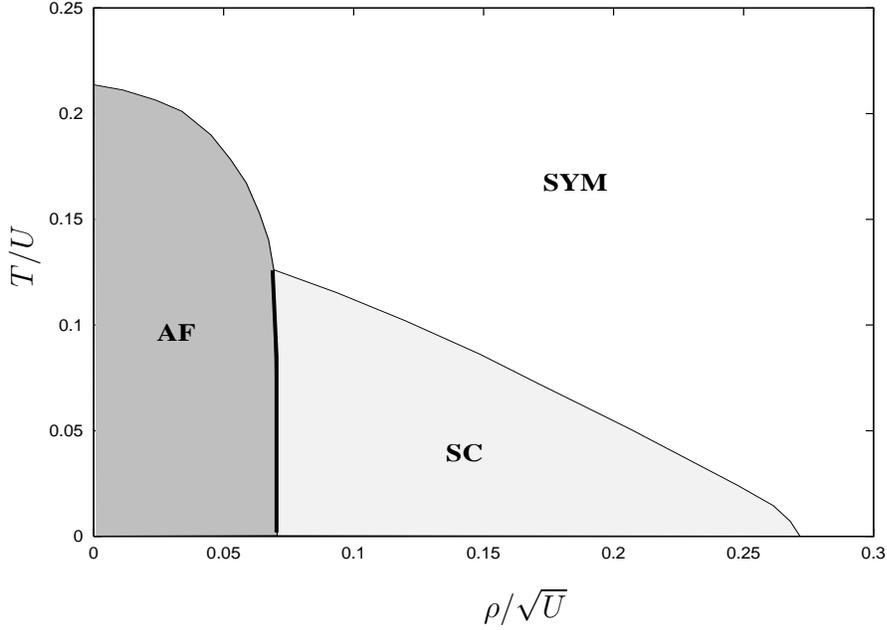,width=150\unitlength,height=100\unitlength}}

\put(80,-5){$\rho/\sqrt{U}$}
\put(5,50){\begin{rotate}{90}$T/U$\end{rotate}}
\end{picture} 

\caption[The $T-\rho$ phase diagram for $h_\rho=h_\alpha=h_\delta=\sqrt{10U}$ with symmetric (SYM), antiferromagnetic (AF) and
superconducting phase (SC). In the region marked by the bold line the phase transition into the antiferromagnetic phase is of first order; all other phase transitions are of second order. ]
{The $T-\rho$ phase diagram for $h_\rho=h_\alpha=h_\delta=\sqrt{10U}$ with symmetric (SYM), antiferromagnetic (AF) and
superconducting phase (SC). In the region marked by the bold line the phase transition into the antiferromagnetic phase is of first order; all other phase transitions are of second order. }
\label{fig:h1}
\end{figure}


\begin{figure}[h]
\setlength{\unitlength}{0.8mm}

\begin{picture}(200,110)(0,-5)

\put(0,0){\epsfig{file=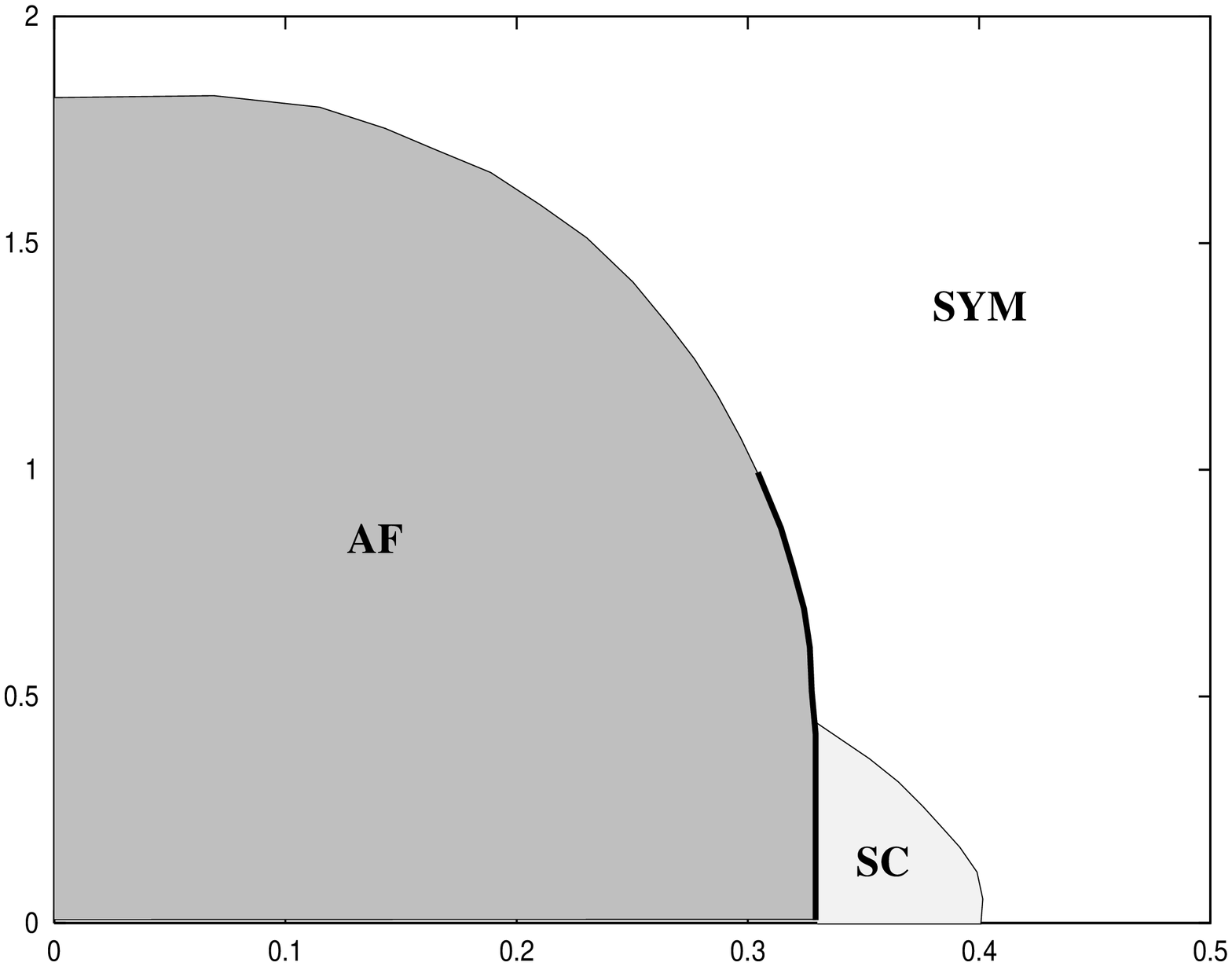,width=150\unitlength,height=100\unitlength}}

\put(80,-5){$\rho/\sqrt{U}$}
\put(5,50){\begin{rotate}{90}$T/U$\end{rotate}}
\end{picture} 

\caption[The $T-\rho$ phase diagram for $h_\rho=h_\alpha=h_\delta=2\sqrt{10U}$ with symmetric (SYM), antiferromagnetic (AF) and
superconducting phase (SC). In the region marked by the bold line the phase transition into the antiferromagnetic phase is of first order; all other phase transitions are of second order.]
{The $T-\rho$ phase diagram for $h_\rho=h_\alpha=h_\delta=2\sqrt{10U}$ with symmetric (SYM), antiferromagnetic (AF) and
superconducting phase (SC). In the region marked by the bold line the phase transition into the antiferromagnetic phase is of first order; all other phase transitions are of second order.}
\label{fig:h2}
\end{figure}

\begin{figure}
\setlength{\unitlength}{0.8mm}

\begin{picture}(200,110)(0,-5)

\put(0,0){\epsfig{file=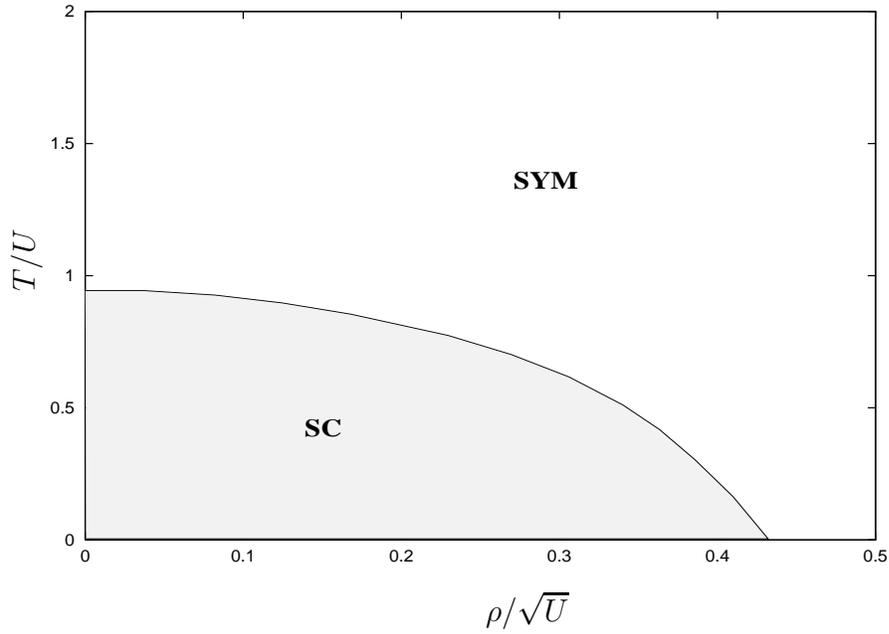,width=150\unitlength,height=100\unitlength}}

\put(80,-5){$\rho/\sqrt{U}$}
\put(5,50){\begin{rotate}{90}$T/U$\end{rotate}}
\end{picture} 

\caption[The $T-\rho$ phase diagram for $h_\alpha=\sqrt{10U}$, $h_\rho=h_\delta=2\sqrt{10U}$ with symmetric (SYM) and
superconducting phase (SC). The phase transitions is of second order.]
{The $T-\rho$ phase diagram for $h_\alpha=\sqrt{10U}$, $h_\rho=h_\delta=2\sqrt{10U}$ with symmetric (SYM) and
superconducting phase (SC). The phase transitions is of second order.}
\label{fig:h12}
\end{figure}

\begin{figure}
\setlength{\unitlength}{0.8mm}

\begin{picture}(200,110)(0,-5)

\put(0,0){\epsfig{file=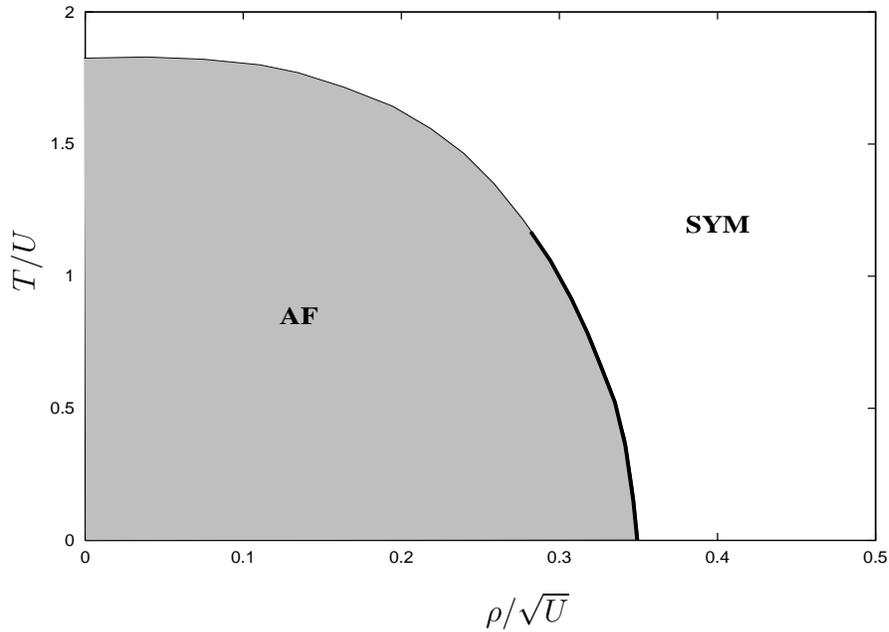,width=150\unitlength,height=100\unitlength}}

\put(80,-5){$\rho/\sqrt{U}$}
\put(5,50){\begin{rotate}{90}$T/U$\end{rotate}}
\end{picture} 

\caption[The $T-\rho$ phase diagram for $h_\delta=\sqrt{10U}$, $h_\rho=h_\alpha=2\sqrt{10U}$ with symmetric (SYM) and
antiferromagnetic phase (AF). In the region marked by the bold line the phase transition into the antiferromagnetic phase is of first order, otherwise of second order. ]
{The $T-\rho$ phase diagram for $h_\delta=\sqrt{10U}$, $h_\rho=h_\alpha=2\sqrt{10U}$ with symmetric (SYM) and
antiferromagnetic phase (AF). In the region marked by the bold line the phase transition into the antiferromagnetic phase is of first order, otherwise of second order. }
\label{fig:h21}
\end{figure}

For vanishing sources $j_a, j_d$ the minimum of $U_0$ obeys the ``field equations''
\begin{eqnarray}
\label{fieldeq.a}
\frac{\partial U_0}{\partial \vec{a}} &=& 2 \vec a \frac{\partial U_0}{\partial \alpha} 
  = 2 \vec{a} \Big(2\Lambda^2 - \frac{1}{2} h_a T\sum_n \int\frac{d^2q}{(2\pi)^2} \\ 
&&\hspace{2.5cm}\mathrm{tr}_8 \big\{ (h_\rho \rho\alpha^{-1/2} A_3\otimes \tau_3 +h_a) 
  \bar P^{-2}_\rho (\alpha,\delta) \big\} \Big) = 0\nonumber \\
\label{fieldeq.d}
\frac{\partial U_0}{\partial d^*} &=& d \frac{\partial U_0}{\partial \delta} 
  = d \Big(4\Lambda^2 - \frac{1}{2} h_d^2 T\sum_n \int\frac{d^2q}{(2\pi)^2} \\ 
&&\hspace{2.5cm}\mathrm{tr}_8 \{[\cos(aq_x)-\cos(aq_y)B_0]^2 \bar P^{-2}_\rho (\alpha,\delta) \}\Big) = 0\nonumber
\end{eqnarray}
with
\begin{eqnarray}
\bar P^2_\rho (\alpha,\delta) &=&  
  P^2_\rho + 2h_\rho h_a \rho \sqrt{\alpha} A_3\otimes\tau_3 \nonumber\\ 
&& {}+ h_a^2\alpha + h_d^2 \delta \left[\cos(aq_x)-\cos(aq_y)B_0\right]^2.
\end{eqnarray}
One always has the symmetric solution $\vec a =0,\, \delta=0$ which corresponds to a local minimum 
if $\bar{M}_a^2>0,\, \bar{M}_d^2>0$ and to a maximum or saddlepoint otherwise. Consider next 
solutions with $\vec a =0,\, \delta\not=0$ which require
\begin{eqnarray}
  \label{sym.brk}
-h_d^4 \, \delta \, T\sum_n \int\frac{d^2q}{(2\pi)^2}\mathrm{tr}_4
  \Big\{ \left[\cos(aq_x)-\cos(aq_y)B_0\right]^4 P_\rho^{-2} \nonumber\\ 
\cdot \left(P_\rho^2 + h_d^2 \delta 
  \left[\cos(aq_x)-\cos(aq_y)B_0\right]^2\right)^{-1} \Big\} \nonumber\\
= 4\Lambda^2 + \Delta M_d^2 &=& \bar M_d^2.
\end{eqnarray}
Solutions with $\delta >0$ exist only for $\bar M_d^2 <0$ and $\delta$ vanishes as the mass term 
$\bar M_d^2$ approaches zero from below. One concludes that the transition from the symmetric 
phase $(\vec a =0, \; \delta=0)$ to a possible superconducting phase without antiferromagnetism
$(\vec a =0, \; \delta>0)$ is of second order. 

We have analyzed the phase diagram for different Yukawa couplings numerically. 
Due to the free parameters $\lambda_i$ in eq. (\ref{14}) the Yukawa couplings are largely undetermined. They only have to obey the inequalities $h_\rho^2 > 0,\; h_d^2 > 0,\; h_a^2 > 0,\; h_a^2 > \pi^2 U/3 + h_\rho^2/3 - 2 h_d^2/3.$ For example, $\lambda_1=\lambda_3=1/2,\; \lambda_2=1$ leads to $h_d^2=h_\rho^2=h_a^2=\pi^2 U/2$. 
Because of our meanfield approximation, the partition function becomes dependent on the particular choice of the parameters $\lambda_i$. We investigate the cases $h_\rho = h_\alpha = h_\delta =\sqrt{10U}$ (fig.~\ref{fig:h1}), 
$h_\rho = h_\alpha = h_\delta =2\sqrt{10U}$ (fig.~\ref{fig:h2}), $h_\alpha = \sqrt{10U}$, $h_\rho=h_\delta =2\sqrt{10U}$ (fig.~\ref{fig:h12}), and $h_\delta = \sqrt{10U}$, $h_\rho=h_\alpha =2\sqrt{10U}$ (fig.~\ref{fig:h21}). We choose $t/U=1$  and investigate the 
phase diagram in the $\rho/\sqrt{U}$-$T/U$-plane. Expressed in the variables $t/U$, $T/U$, $\rho/\sqrt{U}$, our results do not depend on $U$ and the lattice distance $a$, as discussed in the beginning.
As we increase all three Yukawa couplings simultaneously, the antiferromagnetic phase dominates over the superconducting phase (compare figs. \ref{fig:h1} and \ref{fig:h2}). An interesting result of the mean field  analysis is the appearance of a phase transition of first order into the antiferromagnetic phase for small $T/U$ and  for high values of $\rho/\sqrt{U}$. The phase transition between the symmetric and the superconducting phase remains of second order. Both results were anticipated when examining the above formulae analytically. If we increase $h_d/\sqrt{U}$ compared to $h_a/\sqrt{U}$ the superconductivity phase dominates for low $T/U$, whereas in the opposite case it is the antiferromagnetic phase. This is illustrated in figures \ref{fig:h12} and \ref{fig:h21}.

We note that for negative $t$ our results apply if the antiferromagnetic condensate $\vec a$ is replaced by the ferromagnetic condensate $\vec m$. Furthermore, small disturbances can easily be taken into account by source terms. For example, an interaction between spin and angular momentum will explicitely break the continuous $SU(2)$ invariance and typically amount to a source term $l_{\vec a}$ or $l_{\vec m}$.

In conclusion, the mean field approximation for the colored Hubbard model can give a qualitatively reasonable picture of the phases in high $T_c$ superconductors. On the other hand, the shortcomings of this approximation are also apparent from the figures. All phase diagrams in figs. \ref{fig:h1}, \ref{fig:h2}, \ref{fig:h12} and \ref{fig:h21} correspond to different mean field approximations for the same model. It is impossible to resolve this ambiguity within the mean field approximation without additional input on the selection of the Yukawa couplings. The reason is the neglect of fluctuations of the bosonic fields. Only if these are included, the different equivalent choices of the Yukawa couplings should lead to the same physical results. The differences between the figures reveal the importance of the neglected bosonic fluctuations, at least for some choices of the Yukawa couplings\footnote{It is conceivable that an ''optimal choice'' of the Yukawa couplings minimizes the impact of the bosonic fluctuations.}.

The inclusion of the bosonic fluctuations is a complex problem which can be attacked by means of nonperturbative renormalization group equations \cite{fluss}. Studies for similar QCD-motivated models of fermions with Yukawa coupling to scalars have already been carried out successfully \cite{berg}. One of the dominant effects will be the scale dependence of the Yukawa couplings. It is conceivable that this running is dominated by partial infrared fixed points for ratios of Yukawa couplings. For large couplings, as relevant here, such partial fixed points would be approached fast. In this case the ``memory'' of the initial choice of Yukawa couplings could be erased rapidly and unambiguous physical predictions become possible. 

A second important ingredient is the appearance of Goldstone bosons for $\left<\vec a\right> \neq 0$ or $\left<d\right> \neq 0$, corresponding to flat directions in the effective potential (\ref{25}). For a superconducting condensate $\left<d\right>$ the $U(1)$-symmetry would be spontaneously broken and the question arises if this is self consistent. For a large correlation length $\xi$, i.e. $\xi T\gg 1$, one expects that the dominant fluctuations near a second order phase transition are well described by an effective dimensional reduction to two dimensional classical statistics. The Mermin--Wagner theorem then suggests that the Goldstone boson fluctuations prevent a continuous symmetry from being spontaneously broken. In the case of a $U(1)$-symmetry the natural solution to this puzzle is a second order phase transition of the Kosterlitz--Touless type: only a renormalized expectation value differs from zero, whereas the wave function renormalization will lead to a vanishing expectation value for the unrenormalized scalar field \cite{graeter}. This reconciles the Mermin--Wagner theorem with the existence of Goldstone bosons and superconductivity in presence of electromagnetic fields. 

For a possible ``antiferromagnetic phase'' the nonabelian interactions between the Goldstone bosons of the effective two dimensional model have a tendency to push the minimum of $U_0$ towards $\alpha=0$ and to make $\partial U_0/\partial\alpha$ positive \cite{fluss}. If only the nonabelian Goldstone bosons are present in the effective long distance model their fluctuations would destroy the nontrivial minimum of the potential. One may therefore speculate about a new type of low temperature phase, which is characterized by the presence of massless Goldstone bosons as well as massless fermions. Alternatively no true antiferromagnetic phase with Goldstone bosons may occur. For all practical purposes the physics nevertheless will look qualitatively similar to the phase transition in the mean field approximation: the effects from Goldstone fluctuations are only logarithmic in ratios of mass scales and would be cut off by a small $SU(2)$-breaking disturbance inducing a mass term for them. Simple scale considerations suggest that the first
order transitions to the antiferromagnetic phase are probably not affected substantially by the Goldstone fluctuations, 
except for the endpoints. Particularly interesting is the triple point in fig. \ref{fig:h1} where the three phases meet. By continuity of the second order lines one expects five massless scalar
excitations at this point. 

We emphasize that quite generally the possible second order phase transitions between the symmetric and some other phase belong to new interesting universality classes. Long range fermion fluctuations without a gap are present in the symmetric phase and therefore also at the phase transition. They influence the critical exponents and other universal properties.

We hope that our formulation of the colored Hubbard model will be a good starting point for a quantitative renormalization group study of all these interesting questions.




\begin{thebibliography}{99}

\bibitem{hubbard}J. Hubbard, Proc. Roy. Soc. (London) \textbf{A 276}{, 238,
(1963); J. Kanamori, Prog. Theor. Phys.} \textbf{{30}}{,
275, (1963); M. C. Gutzwiller, Phys. Rev. Lett.} \textbf{{10}}{,
159, (1963)}
\bibitem{supra2d} D. J. Scalapino, Phys. Rep. {\bf 250}, 329, (1995)
\bibitem{RG}K. G. Wilson, Phys. Rev. {\bf B4}, 3174, (1971); 3184;
K. G. Wilson, I. G. Kogut, Phys. Rep. {\bf 12}, 75, (1974);
F. Wegner, A. Houghton, Phys. Rev. {\bf A8}, 401, (1973);
F. Wegner in {\em Phase Transitions and Critical Phenomena}, vol. 6, eds. C. Domb and M. S. Greene,
Academic Press (1976);
J. Polchinski, Nucl. Phys. {\bf B231}, 269, (1984)
\bibitem{Sal97} M.~Salmhofer, Commun. Math. Phys. {\bf 194}, 249, (1998)
\bibitem{Met99} C.J.~Halboth, W.~Metzner, cond-mat/9908471
\bibitem{negele}J. W. Negele und H. Orland, \emph{Quantum Many-Particle Systems}, Addison-Wesley,
Redwood City (1988)
\bibitem{fluss} J. Berges, N. Tetradis, C. Wetterich, hep-ph/0005122; C. Wetterich, Phys. Lett. {\bf 301B}, 90, (1993);
Z. Phys. {\bf C48}, 693, (1990); \textbf{{C 57}}{, 451, (1993);}\textbf{{C 60}}{, 461, (1993)}
\bibitem{berg} J. Berges, D.-U. Jungnickel, C. Wetterich, Phys. Rev. {\bf D59}, 034010, (1999);
Eur. Phys. J. {\bf C13}, 323, (2000)
\bibitem{graeter} M. Gr\"ater, C. Wetterich, Phys. Rev. Lett. {\bf 75}, 378, (1995)

\end{thebibliography}
\end{document}